# Impossible Galaxies, the Hubble Tension and the Ricci soliton miracle.

**Stuart Marongwe** [1*]   **Stuart Kauffman** [2]

1. Physics Department, University of Botswana, 4775 Notwane Rd, Gaborone, Botswana stuartmarongwe@gmail.com
2. University of Pennsylvania (emeritus) stukauffman@gmail.com

*corresponding author

**Abstract:** The standard model of cosmology has recently begun to show signs of internal inconsistencies under the relentless onslaught of precision data from the James Webb Space Telescope (JWST), the Hubble Space Telescope (HST) and other space-based observation platforms as well as from those on the ground. The most statistically significant tension, the so-called Hubble Tension, currently stands at between 4-$\sigma$ to 6-$\sigma$ statistical significance. The JWST high resolution images have revealed mature and sometimes quenched galaxies at high redshifts (z>6) which standard cosmological models have failed to both predict and explain. This is the impossible galaxies problem. These short comings of the standard models of cosmology have led to a plethora of new models attempting to explain the inconsistencies between theory and observations. Thus far, none of them adequately address these tensions. In this article, we demonstrate that by including a Ricci soliton into Einstein's field equations, both tensions can be adequately resolved. The Ricci soliton is sourced from the gravitational energy.

## 1. Introduction

Cosmology is currently in a state of crisis as the new era of precision cosmology and high-resolution imaging data begins to reveal inconsistences with the standard models of galactic and cosmic evolution. These inconsistences question the fundamental aspects of how the cosmos works and have generated vigorous debate within the astrophysical community. The debates revolve around questioning the observational methods and the foundational assumptions in the current standard models of cosmology. Resolving these issues will put cosmology on a firmer footing and possibly reveal new physical laws and exotic particles at work in the cosmos.

### 1.1 Hubble Tension

Recent high-resolution near-infrared imaging data from JWST [1-8] and observations made by the HST [9-12], European Space Agency's (ESA) Planck satellite [13-16], combined with other ground-based observations [17-19] challenge the foundations of the current standard model of cosmology, the Lambda Cold Dark Matter (LCDM) model. In particular, there is a disconcerting disagreement between measurements of the Hubble constant in the nearby universe compared with those of the early universe using a variety of observational techniques. For instance, Cosmic Microwave Background (CMB) data from the ESA Planck mission yields $H_0 = 67.2 \pm 1.2$ km/s/Mpc for the early universe observations of the Hubble constant. This value is computed within the framework of the LCDM model. The Hubble constant measurement for the nearby universe with a very low uncertainty (1.9 %) was made by A. Reiss and the SHoES (Supernovae $H_0$ for the Equation of State) team. They used the HST to collect data from 70 Cepheid variables in the Large Magellanic Cloud. The ShoES team measured $H_0 = 73.04.8 \pm 1.04$ km/s/Mpc for the nearby universe. This represents an approximately 10% difference in the measured value at 5-$\sigma$ statistical significance (Fig. 1). In the LCDM model both measurements should agree. This implies that this discrepancy hints at some yet unknown physics or at least that some basic assumptions in the LCDM model have to be revised. W.Freedman using the Gaia Early Data Release 3, employed the Tip of the Red Giant Branch (TRGB) to obtain a value of $H_0 = 69.8 \pm 0.6$ km/s/Mpc. These TRGB results are consistent to within 2 $\sigma$ with the SHoES and Spitzer plus HST Key Project Cepheid calibrations.

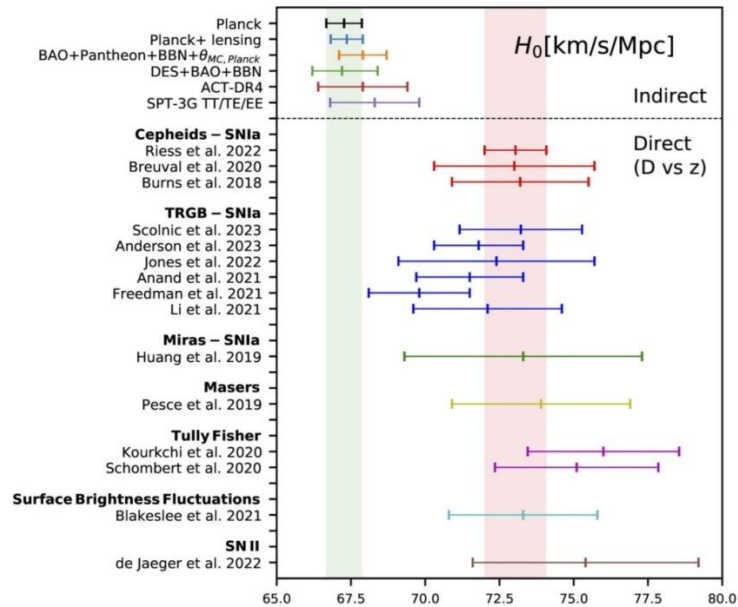

**Fig. 1.** The Hubble constant as measured by different teams. The early universe measurements are depicted by the dotted line and below are the local universe measurements yielding significantly different results, at 4-σ to 6-σ statistical significance. Image credit Di Valentino, E. , 2021, MNRAS 502, 2065

## 1.2 Proposed Solutions to the Hubble Tension

Suggested remedies to the Hubble tension are manifold. One method to resolve the disagreements between conflicting data is to make measurements that are both precise and accurate. This method includes taking measurements of stars from much distance galaxies as well as the use of gravitational waves to improve the quality of the data. However, the ShoES team and others have demonstrated high precision and consistency of the data which currently stands at 4-$\sigma$ to 6-$\sigma$ statistical significance for the existence of the disagreement. At such levels of precision and accuracy, these new techniques might not relieve the Hubble tension. Proposed theoretical models to explain the discrepancies run in the range of a few hundreds and counting  In [20] These models have been classified in the following categories: early dark energy [21-23], late dark energy [24-27], dark energy models with 6 degrees of freedom and their extensions [28-29], models with extra relativistic degrees of freedom [30-33], models with extra interactions [34-37], unified cosmologies 38-40], modified gravity[41-43], inflationary models [44-48], modified recombination history and physics of the critical phenomena [49-50]. Of these models, solutions involving early or dynamical dark energy, neutrino interactions, interacting cosmologies, primordial magnetic fields and modified gravity currently provide the best options thus far until a better alternative comes along.

## 1.3 The Impossible Galaxies Problem

High-resolution near-infrared imaging data from JWST has revealed, quite unexpectedly, high redshift galaxies at the cosmic dawn (Fig. 2) that have disks and bulges similar to those of mature galaxies that have undergone long periods of evolution over cosmic time [51-55]. Furthermore, some of these galaxies are observed to be highly compact, massive and in a state of rapid star formation [56-57]. In the standard models of galaxy formation, galaxies begin their existence as miniscule density fluctuations in which the over dense regions collapse into viralized protogalaxies. These protogalaxies eventually assemble gas and dust into stars and black holes and eventually evolve over cosmic time into spirals and barred spiral galaxies.  The standard models of galaxy formation predict that disk galaxies are too fragile to exist in the early universe due to the high frequency of galactic collisions which destroys their delicate shapes. However, JWST observations reveal that the disk galaxies are ten times more prevalent in the early universe than what standard models predict [58]. This glaring inconsistency completely overturns models of galaxy formation and evolution requiring new logically consistent models which possibly employ new physical concepts.

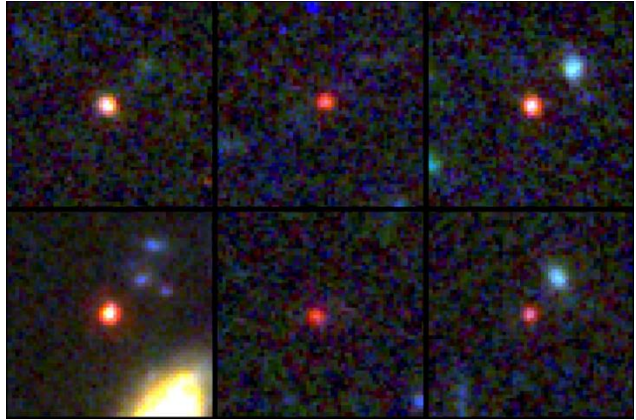

**Fig 2.** Images of six candidate high redshift massive galaxies z>11. Image credit: NASA,ESA,CSA, I.Labbe (Swinburne University of Technology). Image processing G.Brammer (Niels Bohr Institute's Cosmic Dawn Center at the University of Copenhagen)

**1.4 Suggested Solutions**

The main seemingly impassible obstacle in the formation of mature massive early galaxies is the implausible short timescale for star formation. It demands implausible physics that require a much higher fraction of baryons (100%) being converted into stars instantly upon dark matter halo virialization. To remedy the situation, some have suggested identifying flaws in the standard model of cosmology. These include:

1. Failed template fitting or redshift determination. That is, if measurements of star formation rates are wrong it is likely because of the assumption that templates derived from lower redshifts can be used at higher redshifts($z>6$)
2. New clustering physics is explored in which dark matter halos collapse earlier than allowed by current models. This is plausible in the presence of dark energy with $w > -1$. However, the ESA Planck satellite measurement constrain $w \sim -0.95$.
3. Early star formation. This solution considers main sequence star formation occurring much earlier than the initial collapse of halos. However, these ideas are constrained by observations of star formation at lower redshifts as well as the problem of cooling to form stars at low metallicities.

## 2.0 Methodological approach

Our methodological approach to resolve the Hubble tension and the impossible galaxy problem is as follows:

- We apply a combination of analytical and theoretical methods.
- We use observational constraints to guide the theoretical development.
- We employ case studies to demonstrate practical applications.

### 2.1 The Ricci soliton hypothesis

To address the issues raised in the preceding sections we consider a modification of General Relativity, on which the LCDM model is based. The current crisis in cosmology might be a consequence of the absence of a conceptual understanding of the nature of dark energy and dark matter, the very namesake of the model. In Ref. [59] it is demonstrated that dark matter is a Ricci soliton from which the evolving baryonic Tully Fisher relation and Milgrom's acceleration constant are derived from first principles. In the article, a theory of gravity based on two other flavors of the Mach Principle is proposed. These are

1. Space-time cannot exist without matter.
2. Space-time is a compact or a closed system.

To reconcile GR with Mach's principle it is the proposed that the gravitational energy tensor $\tau_{\mu\nu}$ of baryonic matter is associated with a compact 4D Einstein manifold or a trivial Ricci soliton via the expression $\kappa\tau_{\mu\nu} = R_{\mu\nu} - \frac{1}{2}Rg_{\mu\nu} = kg_{\mu\nu}$ and the Ricci tensor of the compact manifold is expressed in terms of the gravitational energy-momentum tensor field $R_{\mu\nu} = \kappa\left(\tau_{\mu\nu} - \frac{1}{2}\tau g_{\mu\nu}\right) = kg_{\mu\nu}$. Here $\kappa$ is the Einstein gravitational constant $k$ is a constant proportional to the cosmological constant. These expressions can only be self-consistent if and only if the traces are $\tau = 0$ and $R = 0$. Here, the compact gravitational stress momentum tensor field $\left(\tau_{\mu\nu} - \frac{1}{2}\tau g_{\mu\nu}\right)$ is considered as a barotropic fluid. By being barotropic, it satisfies the condition of being homogeneous and isotropic in the volume it occupies. Einstein's field equations (EFE) are then modified such that they include a matter generated 4-D compact Einstein manifold and thus expressing them as follows:

$$G_{\mu\nu} - kg_{\mu\nu} + \Lambda g_{\mu\nu} = \kappa(T_{\mu\nu} + \tau_{\mu\nu}) \tag{1}$$

The first term on the left expresses the distortion of space-time by the presence of matter, the second term is a compact Einstein manifold associated with the gravitational

energy and the third term is a contribution by the cosmological constant. We consider the harmonic condition:

$$G_{\mu\nu} - kg_{\mu\nu} = \kappa(T_{\mu\nu} + \tau_{\mu\nu}) - \Lambda g_{\mu\nu} = 0 \tag{2}$$

Such a condition occurs when the gravitational energy density is identical to that of DE. These conditions imply that space-time is covariantly flat and that the gravitional field is self-gravitating.

$$G_{\mu\nu} = kg_{\mu\nu} \tag{3}$$

The harmonic condition implies that Eqn.(3) can be expressed as $\Box \bar{h}_{\mu\nu} = k\bar{h}_{\mu\nu}$ or as system of spherical bound gravitational waves. Clearly, when the harmonic condition is satisfied, the waves have a light like metric implying that the four vector $x^\mu$ exists in a pure state and the waves exhibit E-mode polarity. When Eqn.(2) is positive, the components of the four vector $x^\mu$ become mixed and the metric is time-like. The mixing of the components manifests as gravity implying that Einstein's field equations should be accorded the intrinsic geometric interpretation of curved world lines in flat spacetime. The gravitational waves under this condition have a B-mode polarity, are spherically bound and travel at less than the speed of light. When Eqn.(2) is negative, the gravitational waves become evanescent and only vanishingly low frequency waves can escape the furthest beyond the surface of the confining sphere.

We express the constant $k$ as $k = \alpha^2(t)\Lambda = 3\left(\frac{\alpha(t)H_0}{c}\right)^2$ where $\alpha H_0$ is the expansion rate of the Ricci soliton of reduced radius $r = \frac{c}{2\pi\alpha H_0}$. This implies a rotational velocity on the surface of the soliton of $v = H_0 r = \frac{c}{2\pi\alpha}$. This circular motion on the surface of the Ricci soliton arises from the traceless condition $\tau = 0$. This condition demands marginally stable or zero energy orbits on the surface in which the kinetic energy is equal to the gravitational potential energy.

The harmonic condition of Eqn.(2) demands that Eqn.(1) must satisfy a metric solution of the form

$$-c^2 dt^2 = -c^2 dt^2 + \alpha(t)^2 d\Sigma^2 \tag{4}$$

where $d\Sigma^2 = \frac{dr^2}{1-kr^2} + r^2(d\theta^2 + \sin^2\theta d\phi^2)$ in reduced-circumference polar coordinates.

Eqn.(1) can be expressed as

$$G_{\mu\nu} - \alpha^2(t)\Lambda g_{\mu\nu} + \Lambda g_{\mu\nu} = \kappa(T_{\mu\nu} + \tau_{\mu\nu}) \equiv G_{\mu\nu} + \Lambda' g_{\mu\nu} = \kappa T_{\mu\nu}' \tag{5}$$

where $\Lambda' = (1 - \alpha^2(t))\Lambda$ is the effective cosmological constant and $T_{\mu\nu}' = T_{\mu\nu} + \tau_{\mu\nu}$ is the energy-momentum complex. At large radii from the center of baryonic mass, $T_{\mu\nu}$ becomes vanishingly small and the gravitational energy momentum tensor is dominant. If the metric coefficents of Eqn.(4) are processed in equation (5) we obtain the following analytical solutions given that the gravitational energy-momentum tensor is isotropic and homogeneous:

$$\left(\frac{\dot{\alpha}}{\alpha}\right)^2 + \frac{\alpha^2 \Lambda' c^2}{\alpha^2} - \frac{\Lambda' c^2}{3} = \frac{\kappa c^4}{3}\rho \rightarrow \left(\frac{\dot{\alpha}}{\alpha}\right)^2 + \frac{2\Lambda' c^2}{3} = \frac{\kappa c^4}{3}\rho \tag{6}$$

$$2\frac{\ddot{\alpha}}{\alpha} + \left(\frac{\dot{\alpha}}{\alpha}\right)^2 + \frac{\alpha^2 \Lambda' c^2}{\alpha^2} - \Lambda' c^2 = -\kappa c^2 p \rightarrow 2\frac{\ddot{\alpha}}{\alpha} + \left(\frac{\dot{\alpha}}{\alpha}\right)^2 = -\kappa c^2 p \tag{7}$$

$$R = -6\left[\frac{\ddot{\alpha}}{\alpha} + \left(\frac{\dot{\alpha}}{\alpha}\right)^2 + \Lambda' c^2\right] = 0 \tag{8}$$

From Eqn.(8) we obtain $\frac{\ddot{\alpha}}{\alpha} = -\left(\frac{\dot{\alpha}}{\alpha}\right)^2 - \Lambda' c^2$. placing this result into Eqn.(7) we obtain

$$\left(\frac{\dot{\alpha}}{\alpha}\right)^2 = \kappa c^2 p - 2\Lambda' c^2 \tag{9}$$

Substituting (9) in (6) and taking $p = -\rho c^2$ we obtain

$$\Lambda' c^2 = -\kappa c^4 \rho \rightarrow \Lambda' = -\kappa c^2 \rho \tag{10}$$

Given a measured value of the measured effective cosmological constant of $1.1 \times 10^{-52} m^{-2}$ we compute a critical gravitational mass density for the Ricci soliton of $\rho_c = \rho_{DE} = 5.6 \times 10^{-27} kgm^{-3}$.

The observed Hubble parameter is given by the expression $H_o = \sqrt{\kappa c^2 p - 2\Lambda' c^2} = \sqrt{\kappa c^4 \rho}$. Therefore, the Ricci soliton expands exponentially when $p = -\rho c^2$ with a scale factor $\alpha(t) = \alpha_0 e^{H_o t}$. This accelerated expansion occurs when the covariant flatness condition of Eqn.(2) is satisfied. Under such conditions the centripetal acceleration on the surface of the soliton is given by the expression

$$\frac{v^2}{r} = \frac{(H_0 r)^2}{r} = H_0^2 r = \frac{H_0^2 c}{2\pi \alpha H_0} = \frac{H_0 c}{2\pi \alpha} \tag{11}$$

In the comoving reference frame $\alpha = 1$ which implies a scale invariant acceleration

$$a_0 = \frac{H_0 c}{2\pi} \tag{12}$$

Under such conditions the radius is computed from Eqn.(11) as

$$r = \frac{2\pi v^2}{H_0 c} \tag{13}$$

Substituting $r$ in the expression $\frac{GM(r)}{r} = v^2$ we obtain

$$v^4 = \frac{GM(r)H_0 c}{2\pi} \tag{14}$$

This is the baryonic Tully Fisher relation under the condition that gravitational mass is equivalent to inertial mass and $a_0 = \frac{H_0 c}{2\pi} \cong 1.1 \times 10^{-10} m/s^2$ is the empirically observed Milgrom's acceleration constant [60,61]. This theoretically computed value agrees with observations.

Since $r = r_0 e^{H_0 t}$ then the velocity evolves with the scale factor as follows:

$$v^4 = \frac{GM(r)H_0 c}{2\pi} e^{4H_0 t} \tag{15}$$

We also obtain the following equations of galactic and galactic cluster evolution

$$r = \frac{1}{H_0} e^{(H_0 t)} (GM_g(r) \frac{H_0}{2\pi} c)^{\frac{1}{4}} \quad = \frac{v_n}{H_0} \tag{16}$$

$$v = \frac{dr}{dt} = e^{(H_0 t)} (GM_g(r) \frac{H_0}{2\pi} c)^{\frac{1}{4}} \quad = H_0 r \tag{17}$$

$$a = \frac{dv}{dt} = H_0 e^{(H_0 t)} (GM_g(r) \frac{H_0}{2\pi} c)^{\frac{1}{4}} \quad = H_0 v \tag{18}$$

One important result from Eqn.(1) which is also Eqn.(5) is that these equations are divergence free since they are metric compatible. The far-reaching implication of this result is that the inclusion of the Ricci soliton into GR localizes the gravitational energy momentum tensor- a feat GR struggles to achieve without this modification. Furthermore, these equations satisfy Lovelock's theorem. From a quantum gravity perspective, the Ricci soliton can be considered as a compact graviton field in which graviton confinement is achieved via graviton-graviton interactions. In other words, the Ricci soliton is a graviton ball or gravitonium.

The gravitonium state $(\alpha^2(t)\Lambda g_{\mu\nu} = \kappa \tau_{\mu\nu})$, which is the Ricci soliton term of Eqn.(1) represents a bound state of high-energy gravitons that could explain both the apparent weakness of gravity and the nature of dark matter. In this framework, matter (the

$T_{\mu\nu}$ term of Eqn.(1)) emits high-energy gravitons (the $G_{\mu\nu}$ term of Eqn.(1)) which, instead of propagating freely, rapidly form gravitonium bound states through graviton-graviton interactions that become strong near Planck energies (~$10^{19}$eV). The binding energy scales as $E_b \propto \frac{GE_{high}^2}{\lambda}$. These bound states subsequently decay by emitting a soliton consisting of low-energy gravitons $(\Lambda g_{\mu\nu} = \kappa\tau_{\mu\nu} = \kappa\rho_{DE}g_{\mu\nu})$ which is the dark energy term of Eqn.(1), effectively filtering the gravitational interaction to lower energies. The decay lowers the energy state of the gravitonium state which increases its effective diameter and therefore the compact graviton field expands over cosmic time. This mechanism depicted in Fig.3 could naturally explain why gravity appears weak (we only observe the low-energy decay products) while simultaneously providing a dark matter candidate in the form of long-lived gravitonium states that interact purely gravitationally.

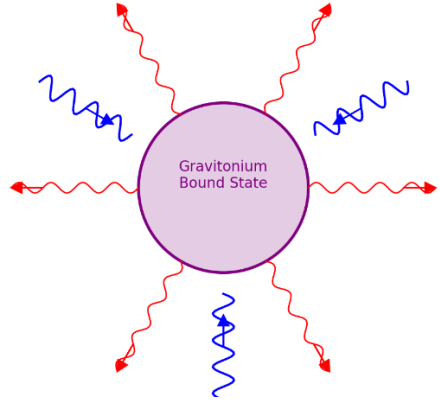

**Fig.3.** The formation and decay mechanism of gravitonium. Wavey blue lines are incoming high energy gravitons that form a marginally stable gravitonium bound state (Dark Matter). The outgoing wavey red lines (Dark Energy) are soft gravitons emitted during the decay of gravitonium.

In rotationally supported galaxies, the visible matter (stars and gas) is distributed in a disk like structure. Therefore the contribution of visible matter to the galaxy rotation curve can be modeled as $v_{visible}(r) = v_{max} \cdot \sqrt{\frac{2}{\pi}} \cdot \arctan\left(\frac{r}{r_s}\right)$ where $r_s$ is the scale radius that determines the extent of the visible matter distribution. Assuming an isothermal sphere approximation with a density profile $\rho(r) \propto \frac{1}{r^2}$ for gravitonium dark matter then we have its contribution to the galaxy rotation curve as

$v_{gravitonium}(r) = v_0 \cdot \sqrt{\ln\left(1 + \frac{r}{r_c}\right)}$. The combined contributions result in a total rotational velocity of $v_{total}(r) = \sqrt{v_{visible}^2(r) + v_{gravitonium}^2(r)}$ which can be plotted in Fig.4

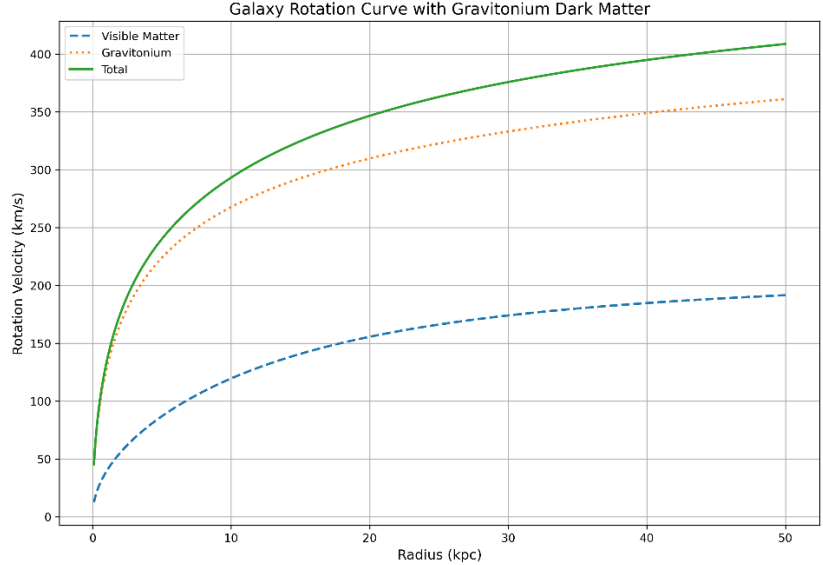

**Fig.4** shows that gravitonium could explain the observed galactic dynamics: The flat rotation curves at large radius is consistent with observations. The approximate mass ratio of gravitonium to visible matter is 3.90 :1 is also consistent with astronomical observations of roughly 5:1 dark matter to visible matter.

However, galaxy and galaxy cluster rotation curves are best modelled by Eqn.(17), which includes the evolutionary properties of galaxy rotation curves over cosmic time. Thus, galaxies are initially compact in the early universe and expand with cosmic time. Lenticulars galaxies, which are older, should exhibit high rotational speeds for less baryonic mass compared to spirals or other younger rotationally supported galaxies. This phenomenon can be detected through an offset between the Baryonic Tully Fisher Relation (BTFR) for spirals compared to that for lenticulars. Eqn.(17) therefore describes an evolving BTFR or eBTFR. The BTFR offset between spirals and lenticulars has been comprehensively studied in [62]. Fig. 5 is a graphical depiction of the empirically observed offset.

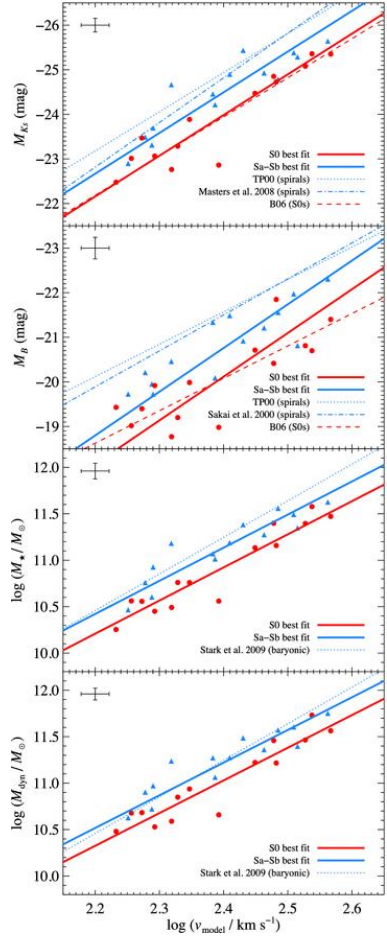

Fig. 5 Tully–Fisher relations for of S0 and spiral galaxies. From top to bottom: total absolute KS-band luminosity, total absolute B-band luminosity, total stellar mass and total dynamical mass. Spiral galaxies are shown as blue triangles and S0 galaxies as red circles. A median error bar is shown in the top left corner of each plot. The best-fitting relations found by inverse regression are shown as solid lines (blue for spirals, red for S0s). The dashed, dotted and dot–dashed lines are a selection of TFRs found by previous authors for different morphological types. Image courtesy Williams.M.J et al, Mon Not R Astron Soc, Volume 409, Issue 4, December 2010, Pages 1330–1346, https://doi.org/10.1111/j.1365-2966.2010.17406.x

Williams *et al* provide strong statistical evidence for a significant offset in the TFRs of S0 and spiral galaxies, with S0s being fainter at a given rotational velocity. They quantify this offset with error estimates and demonstrate that their results are robust against various potential biases and choices of rotational velocity tracers. This lends supports to an evolving baryonic Tully -Fisher relation. A similar analysis with Starburst and gas rich spiral will validate or falsify the eBTFR.

## 2.2. Rapid time evolution in some early galaxies and a resolution of the Hubble Tension

The critical gravitational mass density allows one to compute the volume of the Ricci soliton $V$ and therefore its initial radius $r_0$ from $V = \frac{M_g}{\rho_{DE}}$ yielding

$$r_0 = \left(\frac{3}{4\pi} \cdot \frac{M_g}{\rho_{DE}}\right)^{\frac{1}{3}} \tag{19}$$

If the gravitational mass density $\rho_g > \rho_{DE}$ within the Ricci soliton is concentrated within a sphere of radius $r_g < r_0$ then the critical radius can be computed from

$$r_0 = r_g \left(\frac{\rho_g}{\rho_{DE}}\right)^{\frac{1}{3}} \tag{20}$$

From Eqn.(20) we obtain an expression for the gravitational mass density of

$$\rho_g = \rho_{DE} \left(\frac{r_0}{r_g}\right)^3 \tag{21}$$

The term $r_g$ can be considered as a look-back distance from an observer at the center of the Ricci soliton. The over dense region has an effect on the value of the cosmological constant and the Hubble constant within it via Eqn.(10) which can now be expressed as

$$\Lambda'_g = \kappa c^2 \rho_{DE} \left(\frac{r_0}{r_g}\right)^3 \tag{22}$$

From which we obtain

$$\Lambda'_g = \Lambda' \left(\frac{r_0}{r_g}\right)^3 \text{ and } H_g = H_0 \left(\frac{r_0}{r_g}\right)^{\frac{3}{2}} \tag{23}$$

Since the gravitational mass/energy within the radius $r_g$ is still considered a barotropic fluid then the confinement volume expands exponentially when $p = -\rho c^2$ with a scale factor of

$$\alpha_g(t) = \alpha_{g0} e^{H_g t} \tag{24}$$

The term $H_g t$ can be expressed as $H_0 t_g$ where:

$$t_g = t_{DE} \left(\frac{r_0}{r_g}\right)^{\frac{3}{2}} \tag{25}$$

Here $t_g$ is the time measured by an observer within a zone of gravitational mass density $\rho_g$ and $t_{DE}$ is the time measured by an observer within a zone of gravitation energy density $\rho_{DE}$. Thus, we note rapid time evolution within the over dense region compared to a region at the critical gravitational mass density. This aspect explains why some infant galaxies matured faster than others which were initially less dense. We also note that the equations (23)-(25) might explain the inflationary process during the Big Bang which might render the inflaton field hypothesis obsolete.

The inflationary dynamics of the Ricci soliton is governed by the Ricci flow equation:

$$\frac{\partial g(t)}{\partial t} = -2Ric(g(t)) \quad (26)$$

where $g(t)$ is the metric tensor on the manifold at time $t$ and $Ric(g(t))$ is the Ricci curvature tensor of the metric $g(t)$. Just as heat diffuses in a medium, Ricci flow diffuses or redistributes the curvature of a manifold. Regions with high curvature or high gravitational energy tend to expand, while areas with low curvature contract, leading over time to a more homogeneous distribution of curvature across the manifold as described by Eqn.(2).

## 2.3 Hoag's Object as an example of a typical Ricci soliton

Hoag's Object (PGC 54559, PRC D-51) as depicted in Fig.6, is a ring galaxy type E0 or SA0/a in the constellation of Serpens Caput. The galaxy has a $D_{25}$ isophotal diameter of 45.41 kiloparsecs. A nearly perfect ring of young hot blue stars orbits the older yellow nucleus. The mass estimate by Schweizer *et al* [63] places it at $700 \, GM_*$ which is mostly concentrated in the inner core of diameter 6 arcseconds or $5.3 \pm 0.2$ kpc. There is a gap between the galactic core and the inner radius of the ring with little to no baryonic matter. Brosch *et al* [64] show that the luminous ring lies at the inner edge of a much larger neutral hydrogen ring. The formation history of Hoag's object has been a mystery since its discovery. Most ring galaxies are generally formed by the collision of a small galaxy with a larger disk-shaped galaxy. In this case, there is no sign of a smaller galaxy that would have collided with it.

Given the circular velocity of the ring and its inner radius one can compute the centripetal acceleration of the ring which interestingly satisfies Eqn.(11). Moreover, the inner radius satisfies Eqn.(19). Therefore, the inner radius marks the initial radius of an expanding Ricci soliton which satisfies Brosch *et al*'s observation. The difference in age between the stellar populations is due to rapid time evolution at the core relative to the ring as dictated by Eqn.(25). This also implies that the values of the Hubble constant and the cosmological constant are higher at the core than at the ring as dictated by Eqn.(23). Moreover, the high value of both constants lowers the gravitational potential at the core which favors ultrafast out flows (UFOs) from the core by hot and low molecular mass gases such as hydrogen and helium which then accumulate at the critical radius and beyond to form star nurseries. These UFOs also quickly quench the core and deny it any new star formation.

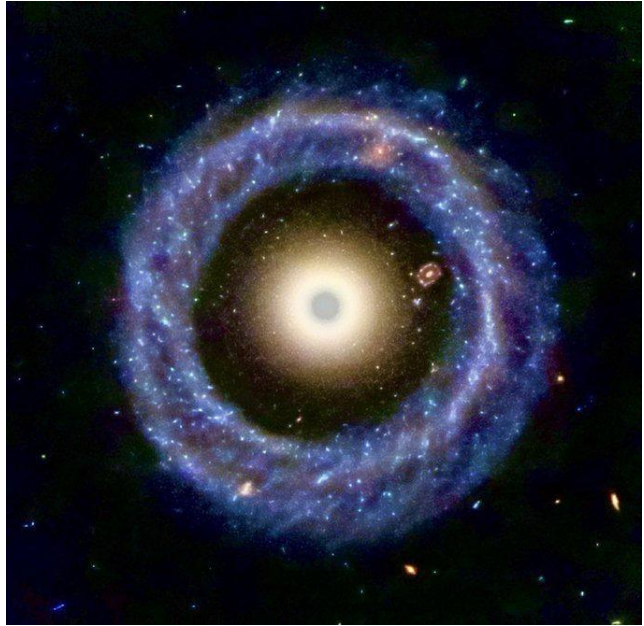

**Fig.6** Hoag's Object is a rare ring galaxy located 600Mlys from Earth characterized by a near perfect ring abundant with young blue hot stars orbiting a galactic core populated by mostly old red giant stars. The Ricci soliton hypothesis explains its enigmatic structure formation and properties which strongly reflects the physics in our galactic neighborhood.

Hoag's Object therefore strongly reflects what is happening in our galactic neighborhood. We are located in an over dense region near the center of a Ricci soliton implying that there is a gap between the core and the initial radius of the Ricci soliton. Mazurenko *et al* [65] have identified that our local galactic neighborhood resides in a dense region of baryonic matter inside the KBC void and their studies reveal that this void is indeed the source of the Hubble tension.

The physical laws governing Hoag's Object are also at work in young compact galaxies in the early universe. These laws are less likely to be pronounced in low baryonic density galaxies. The same laws explain the existence of ultra-diffuse galaxies if their cores are mostly populated by compact stellar objects such as black holes, neutron stars or supermassive black holes. Star formation only takes place after the critical radius just as in Hoag's Object. The more massive the core the larger a galaxy becomes.

Interestingly the bullet cluster shows Ricci solitons shown in blue with compact galactic clusters at their core as depicted in Fig.7a. The diffuse intra galactic gas shown in red is not dense enough to have pronounced effects that reveal an associated Ricci soliton. The galactic cluster CI 0024+17 (ZwCi0024 + 1652) shown in Fig. 7b also shares similarities with Hoag's object with its dense core, a gap and an outer ring. In Fig.7b clusters of Ricci solitons are at the core and along the critical radius of a larger expanding Ricci soliton. In all galactic clusters, the central galaxy must appear older than the surrounding galaxies.

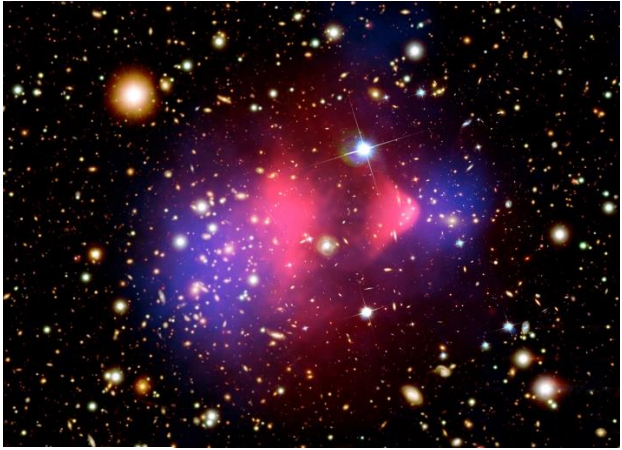
Fig. 7a

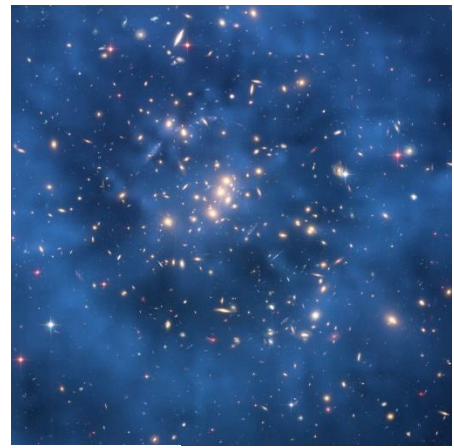
Fig. 7b

**Fig. 7a** is the Bullet Cluster (image credit NASA/CXC/Cfa/M.Markevitch; Optical and lensing map:NASA/STSCL,Megellan/U of Arizona/D.Clowe lensig map) reveals the Ricci soliton through gravitational lensing as well as in **Fig. 7b** in the galactic cluster CI 0024+17 (image credit NASA,ESA,M.J. Lee ,H.Ford of John Hopkins University). In **Fig.7b** clusters of Ricci solitons are at the core and along the critical radius of a larger expanding Ricci soliton. The core galaxies should appear older than the ring galaxies.

**3.0 Observational Tests and Future Prospects**

The Ricci soliton model has several far-reaching implications. For instance, it provides alternative explanation for galactic rotation curves without the need for exotic dark matter particles and naturally explains the baryonic Tully-Fisher relation as well as derives the empirically observed Milgromian acceleration constant from first principles. Moreover, it naturally predicts a varying cosmological constant based on matter density which naturally explains local vs. global differences in expansion rates while simultaneously providing a framework for understanding void dynamics.

The Ricci soliton framework therefore makes several testable predictions:
With regards to structural properties the model shows that there is

- Correlation between core density and evolution rate
- Specific morphological features
- Galaxy/cluster mass-size relations

These structural properties predicted by the model will be better tested in core-ring morphologies resembling Hoag's object.

The model also provides a way of observing distribution functions such as

- Mass function evolution
- Size distribution
- Morphological type fractions

Finally, the model also offers a framework for exploring environmental effects such as

- Density-dependent evolution rates
- Cluster vs. field differences
- Void galaxy properties

Critical tests of this framework require enhanced observations which includes higher resolution imaging, spectroscopic confirmation and environmental studies. Statistical analysis such as large sample studies, multi-wavelength observations and environmental correlations will also offer pathways to falsifying the Ricci soliton model. We suggest further theoretical developments via numerical simulations, semi-analytic models and machine learning applications.

These critical tests will provide a comprehensive context for understanding both the nature of impossible galaxies and potential pathways to resolving the apparent Hubble tension and the evolution of the seemingly impossible galaxies. The Ricci soliton approach offers a promising direction that naturally addresses both the observational discrepancy and related cosmological puzzles.

### 3.1 Key observational predictions

Space-Based facilities such as the JWST deep field observations, Roman wide-field surveys and Euclid complementary data as well as Ground-Based support from the ELT spectroscopic follow-ups, SKA radio observations and LSST time domain will further validate or falsify the Ricci soliton hypothesis. The analysis will include multi-wavelength synthesis, cross-correlation studies and machine learning applications. Short term test (1-5years) using these observational platforms are capable of validating or falsifying core-ring morphology statistics, local expansion rate variations and density-age correlations. Medium-Term Tests (5-10 years) will be capable of validating or falsifying high-z galaxy demographics, cluster evolution patterns and void structure evolution. Finally, Long-Term Tests (>10 years) will validate or falsify cosmic web evolution, dark sector interactions and quantum gravity effects. Interestingly, Baryonic Acoustic Oscillations show the same ring morphologies as Hoag's Object. More data is needed to confirm whether there is a correlation between size of the ring and the mass of the central core as described by Eqn.(19).

### 3.2 Laboratory tests for the Ricci soliton/gravitonium

Laboratory scale experiments to detect the Ricci soliton using gravitational lensing can be performed using an isolated Tritium nucleus, which has a mass of approximately three hydrogen atoms. Inputting the Tritium mass into Eqn.(19) we obtain a value for the gravitonium radius of $r_0 \sim 0.6m$ which is also the impact parameter for gravitational lensing. Given the ingenuity of experimental physicists, a laboratory experiment can be designed and conducted to directly detect gravitonium. The quantum gravity properties of a Ricci soliton or gravitonium have been described in [66,67].

**Discussion and concluding remarks.**

In this article, we have demonstrated that by simply adding a Ricci soliton to GR, the Hubble tension and the impossible galaxies can be explained. The 10% difference between early universe and nearby universe measurements of the Hubble constant are a consequence of the fact that the Milky Way galaxy resides in region of high baryonic matter density within the KBC void. The enigma of mature galaxies in the early universe is resolved by the understanding that those with dense cores experience rapid time evolution than those with less. Furthermore, rapid time evolution increases the power output of stars making the galactic cores appear brighter and hence apparently more massive than they actually are. Mach's principle associates baryonic matter with a compact Einstein manifold via the gravitational stress energy tensor. The Ricci soliton comes with its gravitational energy which initially is equivalent in mass to the baryonic mass content in it. As it expands, its energy/mass increases via self-interaction. the extra gravity manifests as dark matter. Thus, the older a galaxy, the more dark matter content it has. As a final note, GR can satisfactory describe the large-scale gravitational physics in the form $G_{\mu\nu} + \Lambda' g_{\mu\nu} = \kappa T_{\mu\nu}'$. We are of the opinion that if more studies on the far-reaching consequences of including the Ricci soliton in GR are carried out, then cosmology and astrophysics will be on a much firmer footing with less internal inconsistencies.

## Data Availability

Empirical data used in this research can be found in the cited articles and can be obtained from the authors upon reasonable request.

## Acknowledgements

The authors gratefully appreciate the discussions, suggestions and constructive criticism from John C. Mathers, George Ellis and Christian Corda.

## Conflict of interest

We declare no conflict of interest.

## References

1. Castellano,M *et al*, *Early Results from GLASS-JWST. III. Galaxy Candidates at z ~9–15*. 2022 ApJL 938 L15
2. Pontoppidan, K.M *et al*, *The JWST Early Release Observations*,2022 ApJL 936 L14
3. Robertson, B.E, *Galaxy Formation and Reionization: Key Unknowns and Expected Breakthroughs by the James Webb Space Telescope.* Annual Review of Astronomy and Astrophysics Volume 60, 2022
4. Treu. T *et al*, *The GLASS-JWST Early Release Science Program. I. Survey Design and Release Plans*, 2022 ApJ 935 110
5. Carnall,A.C *et al* , *A first look at the SMACS0723 JWST ERO: spectroscopic redshifts, stellar masses, and star-formation histories*, Monthly Notices of the Royal Astronomical Society: Letters, Volume 518, Issue 1, January 2023, Pages L45–L50,
6. Pascale, M. *et al*, *Unscrambling the Lensed Galaxies in JWST Images behind SMACS 0723*,2022 ApJL 938 L6
7. Merlin, E *et al* , *Early Results from GLASS-JWST. II. NIRCam Extragalactic Imaging and Photometric Catalog*,2022 ApJL 938 L14
8. Ferreira, L. *et al*, *Panic! at the Disks: First Rest-frame Optical Observations of Galaxy Structure at z > 3 with JWST in the SMACS 0723 Field*,2022 ApJL 938 L2
9. Reiss,A.G. et al, *Observational evidence from supernovae for an accelerating universe and a cosmological constant*, The Astronomical Journal, 116:1009È1038, 1998 September

10. Riess, A.G. *Using Type Ia Supernova Light Curve Shapes to Measure the Hubble Constant*. 1995 The Astrophysical Journal. 438: L17.
11. Di Valentin,E *et al*, *In the realm of the Hubble tension—a review of solutions* , 2021 Class. Quantum Grav. 38 153001
12. Riess,A.G., Stefano, C., Wenlong ,Y., Lucas, M. M., Dan, S., *Large Magellanic Cloud Cepheid standards provide a 1% foundation for the determination of the Hubble constant and stronger evidence for physics beyond $\Lambda$CDM*. The Astrophysical Journal, 876:85 (13pp), 2019 May 1
13. Planck Collaboration, *Planck 2018 results. V. CMB power spectra and likelihoods,* ,2020, A&A, 641, A5
14. Planck Collaboration, *Planck 2018 results. VI. Cosmological parameters*, 2020, A&A, 641, A6
15. Planck Collaboration, *Planck 2018 results. VII. Isotropy and Statistics of the CMB* 2020, A&A, 641, A7
16. Planck Collaboration, *Planck 2018 results. VIII. Gravitational lensing*,2020, A&A, 641, A8
17. Blanton, M.R *et al*. *Sloan Digital Sky Survey IV: Mapping the Milky Way, Nearby Galaxies, and the Distant Universe*, 2017 AJ 154 28
18. Majewski, S.R et al, *The Apache Point Observatory Galactic Evolution Experiment (APOGEE)*, 2017 AJ 154 94
19. DESI Collaboration *et al.*, *DESI 2024 III: Baryon Acoustic Oscillations from Galaxies and Quasars*, arXiv:2404.03000 [astro-ph.CO]
20. Di Valentino,E *et al*, *In the realm of the Hubble tension—a review of solutions*, 2021 Class. Quantum Grav. 38 153001
21. Caldwell R R, Doran M, Mller C M, Schfer G and Wetterich C , *Early Quintessence in Light of the Wilkinson Microwave Anisotropy Probe*, 2003 Astrophys. J. 591 L75–8
22. Bartelmann M, Doran M and Wetterich C, *Non-linear structure formation in cosmologies with early dark energy*, 2006 Astron. Astrophys. 454 27–36
23. Doran M, Karwan K and Wetterich C, *Observational constraints on the dark energy density evolution*, 2005 J. Cosmol. Astropart. Phys. JCAP11(2005)007
24. Zhao, G. B. *et al.*, *Dynamical dark energy in light of the latest observations* , 2017 Nat. Astron. 1 627–32
25. Di Valentino E, *Crack in the cosmological paradigm*, 2017 Nat. Astron. 1 569–70
26. Bonilla A, Kumar S and Nunes R C, *Measurements of and reconstruction of the dark energy properties from a model-independent joint analysis*, 2021 Eur. Phys. J. C 81 127
27. Wang Y, Pogosian L, Zhao G-B and Zucca A, *Evolution of Dark Energy Reconstructed from the Latest Observations*, 2018 Astrophys. J. 869 L8

28. Hu,W., Crossing the Phantom Divide: *Dark Energy Internal Degrees of Freedom*, Phys. Rev. D 71, 047301 2 February 2005
29. Kucukakca, Y.,Akbarieh, A. R and Ashrafi,S, *Exact solutions in teleparallel dark energy model*, Chinese Journal of Physics Volume 82, April 2023, Pages 47-61
30. Steigman G, Schramm D N and Gunn J E. *Cosmological limits to the number of massive leptons*, 1977 Phys. Lett. B 66 202–4
31. Mangano G, Miele G, Pastor S, Pinto T, Pisanti O and Serpico P D, *Relic neutrino decoupling including flavour oscillations*, 2005 Nucl. Phys. B 729 221–34
32. de Salas P F and Pastor S, *A precision calculation of relic neutrino decoupling*, 2016 J. Cosmol. Astropart. Phys. JCAP07(2016)051
33. Akita K and Yamaguchi M 2020 J. Cosmol. Astropart. Phys. JCAP08(2020)012
34. Chanda, A., Halder, A., Majumdar, A.S. et al. *Late time cosmology in $f(\mathcal{R},\mathcal{G})$ gravity with exponential interactions*. Eur. Phys. J. C 83, 23 (2023).
35. Capozziello,S., De Laurentis,M., *Extended theories of gravity*, Phys. Rep. 509, 167 (2011)
36. Faraoni, V., Capozziello,S., *Beyond Einstein Gravity: A Survey of Gravitational Theories for Cosmology and Astrophysics*, Fundam.Theor. Phys., 170 (2010), (Springer, Dordrecht, 2011)
37. Nojiri, S., Odintsov,S.D. and V.K. Oikonomou, *Modified gravity theories on a nutshell: Inflation, bounce and late-time evolution*, Phys. Rep. 692, 1 (2017)
38. Kamenshchik ,A., Moschella, U. and Pasquier V, *An alternative to quintessence*, 2001 Phys. Lett. B 511 265–8
39. Bilic .N., Tupper ,G. B. and Viollier ,R.D., *Unification of dark matter and dark energy: the inhomogeneous Chaplygin gas* , 2002 Phys. Lett. B 535 17–21
40. Gorini V, Kamenshchik A and Moschella U., *Can the Chaplygin gas be a plausible model for dark energy?*, 2003 Phys. Rev. D 67 063509
41. Capozziello ,S., *Curvature quintessence*, 2002 Int. J. Mod. Phys. D 11 483–91
42. Nojiri, S. and Odintsov, S. D., *Modified Gravity with ln R Terms and Cosmic Acceleration*, 2004 Gen. Relativ. Gravit. 36 1765–80
43. Carloni, S., Dunsby, P. K. S., Capozziello, S. and Troisi, A., *Cosmological dynamics of Rn gravity*, 2005 Class. Quantum Grav. 22 4839–68
44. Guth, A. H. and Pi S-Y, *Fluctuations in the new inflationary universe*,1982 Phys. Rev. Lett. 49 1110–3
45. Starobinsky, A. A., *Dynamics of phase transition in the new inflationary universe scenario and generation of perturbations*, 1982 Phys. Lett. B 117 175–8

46. Linde, A. D., *A new inflationary universe scenario: a possible solution of the horizon, flatness, homogeneity, isotropy and primordial monopole problems,* 1982 Phys. Lett. B 108 389–93

47. Albrecht, A. and Steinhardt, P. J., *Cosmology for grand unified theories with radiatively induced symmetry breaking,* 1982 Phys. Rev. Lett. 48 1220–3

48. Colless, M. *et al.*, *The 2df galaxy redshift survey: spectra and redshifts,* 2001 Mon. Not. R. Astron. Soc. 328 1039

49. Chiang, C. T. and Slosar, A., *Inferences of $H_0$ in presence of a non-standard recombination,* 2018 arXiv:1811.03624

50. Liu, M., Huang, Z., Luo, X., Miao, H., Singh, N. K. and Huang, L., *Can non-standard recombination resolve the Hubble tension?,* 2020 Sci. China Phys. Mech. Astron. 63 290405

51. Rhoads, J.E. *et al*, *Finding Peas in the Early Universe with JWST,* 2023 ApJL 942 L14

52. Wang,T. *et al*, *The true number density of massive galaxies in the early Universe revealed by JWST/MIRI,* arXiv:2403.02399 [astro-ph.GA]

53. Nanayakkara, T., Glazebrook, K., Jacobs, C. et al. *A population of faint, old, and massive quiescent galaxies at $3 < z < 4$ revealed by JWST NIRSpec Spectroscopy.* Sci Rep 14, 3724

54. Whitler, L. *et al*, *On the ages of bright galaxies ~500 Myr after the big bang: insights into star formation activity at $z \gtrsim 15$ with JWST,* Monthly Notices of the Royal Astronomical Society, Volume 519, Issue 1, February 2023

55. Melia, F., *The cosmic timeline implied by the JWST high-redshift galaxies,* Monthly Notices of the Royal Astronomical Society: Letters, Volume 521, Issue 1, May 2023, Pages L85–L89

56. Carnall, A.C. *et al*, *The JWST EXCELS survey: too much, too young, too fast? Ultra-massive quiescent galaxies at $3 < z < 5$,* Monthly Notices of the Royal Astronomical Society, Volume 534, Issue 1, October 2024, Pages 325–348,

57. Johnson, J.L. *et al*, *The first galaxies: signatures of the initial starburst,* Monthly Notices of the Royal Astronomical Society, Volume 399, Issue 1, October 2009, Pages 37–47

58. Robertson, B.E. *et al*, *Morpheus Reveals Distant Disk Galaxy Morphologies with JWST: The First AI/ML Analysis of JWST Images,* 2023 ApJL 942 L42

59. Marongwe, S. and Kauffman, S.A., *Dark Matter as a Ricci Soliton,* arxiv.0907.2492v2[physics.gen-ph]

60. Milgrom, M., *A modification of the Newtonian dynamics — Implications for galaxies,* 1983 Astrophys. J. 270 371–89

61. Milgrom, M., *A modification of the Newtonian dynamics as a possible alternative to the hidden mass hypothesis*, The Astrophysical Journal, 270:365-370, 1983 July 15
62. Williams.M.J et al, Mon Not R Astron Soc, Volume 409, Issue 4, December 2010, Pages 1330–1346, https://doi.org/10.1111/j.1365-2966.2010.17406.x
63. Schweizer, F., Ford, W. K., Jedrzejewski, R., Giovanelli, R., *The Structure and Evolution of Hoag's Object*. 1987 The Astrophysical Journal. 320: 454.
64. Brosch, N., Finkelman, I.; Oosterloo, T., Jozsa, G. and Moiseev , A., *HI in HO: Hoag's Object revisited*. 2013 Monthly Notices of the Royal Astronomical Society. 435 (1): 199–206.
65. Mazurenko,S., Banik,I., Kroupa, P and Haslbauer, M., *A simultaneous solution to the Hubble tension and observed bulk flow within 250 h−1 Mpc, Monthly Notices of the Royal Astronomical Society*, Volume 527, Issue 3, January 2024, Pages 4388–4396
66. Marongwe, S.,*Horizon scale tests of quantum gravity using the event horizon telescope observation*,2023 Int. J. Mod. Phys. D 32 2350047
67. Marongwe, S., *Energy momentum localization in quantum gravity* , 2024 Phys. Scr. 99 025306